\documentclass[final,5p,times,twocolumn]{elsarticle}

\usepackage{latexsym}
\usepackage{amssymb, amsmath, bm, physics}
\usepackage{subcaption}
\usepackage{mathptmx}
\usepackage{flushend}
\usepackage[colorlinks,citecolor=blue,urlcolor=blue,linkcolor=blue]{hyperref}
\usepackage{aas_macros}
\usepackage{orcidlink}

\biboptions{sort&compress}

\begin{document}

\begin{frontmatter}




\title{Quantum Black hole--White hole entangled states}


\author{S. Jalalzadeh\orcidlink{0000-0003-4854-2960}}
\address{Departamento de F\'{i}sica, Universidade Federal de Pernambuco, Recife, PE, 52171-900, Brazil}
\ead{shahram.jalalzadeh@ufpe.br}

\begin{abstract}
We investigate the quantum deformation of the Wheeler--DeWitt equation of a Schwarzchild black hole. Specifically, the quantum deformed black hole is a quantized model constructed from the quantum Heisenberg--Weyl $U_q(h_4)$ group. We show that the event horizon area and the mass are quantized, degenerate, and bounded. The degeneracy of states indicates entangled quantum black hole/white hole states. Accordingly, quantum deformation provides a new framework to examine Einstein--Rosen wormhole solutions. Besides, we obtain the mass, the temperature, and the entropy of the q-deformed quantum Schwarzschild black hole. We find an upper bound on the mass of a black hole/white hole pair. Also, at the quantum deformation level, the entropy of the black hole contains three parts: the usual Bekenstein--Hawking entropy, the logarithmic term, and a Cube of usual black hole entropy.
\end{abstract}


\end{frontmatter}



\section{Introduction}
\label{sec:intro}

Today, there is considerable  observational evidence of the existence of black holes (BH). We can find them commonly in binary systems where the orbital motion of the companion of the BH  is used to estimate the mass of the obscured BH \cite{Ritter1}. Astrophysicists believe that most galaxies have a supermassive BH at their center \cite{Rees}. 
They have discovered at least two classes of BH candidates \cite{Stellar}. In the first class, we have stellar-mass objects in X-ray binary systems while supermassive BHs at the center of normal galaxies belong to the second group. Also, there is a set of observations intimating that BH candidates have indeed an  event horizon \cite{Event}. They have measured gravitational waves, resonating from BHs colliding, and finally, astrophysicists announced they had captured the first-ever image of a BH at the center of galaxy M87 \cite{Aki1,Aki2}.

The theoretical developments of BHs led us to the hypothesis that white holes (WH) or even wormholes may exist. Researchers have paid significantly less engagement to WHs than BHs. This is comprehensible given that conditions in our Universe lead to the production of BHs, but WH generation has never been observed or is predicted to occur in the Universe's history, and If there are wormholes, they are tiny, or like dark matter, they only interact with ordinary matter, which is why we have not seen them before. Our analysis is theoretical rather than empirical astrophysical interest. Our model applies to a BH at the end of the evaporation process, where the BH does not simply disappear: it changes into a WH state with a mass of the order of Planck mass, and then it slowly emits thermal radiation and evaporates, possibly only after a long time. 

In the early eighties of XX century, quantum aspects of BH became an interesting subject of study for theoretical physicists, when Bekenstein postulated that the entropy of a BH is proportional to its horizon area \cite{Bec1} and he argued that the possible eigenvalues of the BH’s horizon area are
\begin{equation}\label{AA1}
A_n=\gamma L_\text{P}^2n,~~~~n=1,2,3,...~,
\end{equation}
where $\gamma$ is a dimensionless constant of  order one and $L_\text{P}$ is the Planck length.
 The pioneer works of Bekenstein initially led to the physics of  BHs with entropy but no concept of temperature. The paradox was resolved when the evaporation of BHs was put forward by Hawking \cite{Hawking1} and after that, it was realized by physicists that there is an intimate connection between horizons and temperature \cite{Fulling1}.
Later works in quantum mechanics of BHs over three past decades have re-derived the results of the above references and extended them in many different directions. In this context, various arguments has been introduced in favor of the area spectrum (\ref{AA1}), include information-theoretic considerations \cite{Area1a,Area1c}, string theoretic arguments \cite{Area2}, periodicity of time \cite{Area3d} and a Hamiltonian quantization of a dust collapse \cite{Area4a}. 

In the above arguments of quantization of area, a BH is the vacuum solution of the Einstein field equations. In the real world, a BH is embedded in the Universe with event horizon $A_\text{Universe}$. As we know, the Schwarzschild event horizon (SEH) of a BH, $R_{SEH}$ can not be bigger than the cosmic event horizon (CEH) $L_\text{CEH}$ of the Universe at a given moment, $A_\text{SEH}\leq A_\text{CEH}$. Therefore, in the presence of a cosmological horizon, the dimension of the Hilbert space of the area given by Eq.(\ref{AA1}) have to be finite and 
we need to put a bound on the maximum value of the natural number $n$ by
$n_\text{Max}\leq \left(\frac{L_\text{CEH}}{L_\text{P}}\right)^2.$
One method to {retrieve the dimension of a Hilbert space into a finite value is through quantum deformation (utilizing quantum groups) of the model, when the deformation parameter is a root of unity} \cite{1994PhLB..331..150B,Area4c}. Historically, quantum groups {have emerged from studies on quantum integrable models, using quantum inverse scattering methods,  which led to the deformation of classical matrix groups and their corresponding Lie algebras} \cite{kulish1983quantum1}. Recently, quantum groups {were found to} play a major role in quantum integrable systems \cite{chari1995guide}, conformal field theory \cite{oh1992conformal}, knot theory \cite{kauffman2007q}, solvable lattice models \cite{foda1994vertex}, topological quantum computations \cite{nayak2008non},  molecular spectroscopy \cite{chang1991q} and quantum gravity \cite{smolin1995linking1,Area4c,Jalalzadeh:2021oxi}. When the deformation parameter is the root of unity, there exist indeed unitary representations. 
Furthermore, this case appears in applications, such as in duality properties of conformal field theory (see for example \cite{alvarez1990duality}) in which 
$q$ appears in the $SU(2)$ Witten--Wess--Zumino (WZW) model with the value $q=e^\frac{i\pi}{k+2}$,  where $k$ is the level of the corresponding Kac--Moody algebra. Another example is the chiral Potts model \cite{date1991generalized} and its generalizations. 

The rest of the paper is as follows. In section II we review the canonical quantization and the entropy of a Schwarzschild BH in the reduced Hamiltonian theory of BHs. In Section III we generalize the quantum mechanics of  Schwarzschild BHs to noncommutative phase space. We consider the q-deformed BH for $q$ being a root
of unity. We obtain the event horizon and the mass of a BH and after that, we extend the results of the second section to obtain the entropy of a q-deformed BH. Section IV contains a brief summary and a discussion. 
For the remainder of this paper, we shall work in natural units, $\hbar=c=k_\text{B}=1$.

\section{Quantum Mechanics of the Schwarzschild Black Hole}
Let us summarize
a few essential results concerning the quantization and the entropy of the Schwarzschild BH.
The spherical symmetric 
ADM line element is
\begin{multline}\label{1-1}
ds^2=-N(r,t)^2dt^2+\\\Lambda(r,t)^2(dr+N^r(r,t)dt)^2+R(r,t)^2d\Omega^2,
\end{multline}
where $d\Omega^2$ is the line element for the unit two sphere $S^2$. We 
follow Kucha\v r's  fall-off conditions \cite{Kuchar1},
which guarantees that the coordinates $r$ and $t$ are extended to the Kruskal manifold, $-\infty< r,t<\infty$ and that the  spacetime is asymptotically flat as well. 
Then, the Hamiltonian form of the Einstein--Hilbert action functional will be
\begin{multline}\label{1-2}
 S=\displaystyle\int dt\int_{\Sigma_r}\Big\{ \Pi_\Lambda \dot\Lambda +\Pi_R\dot R-NH-N^rH_r\Big\}dr -\\\displaystyle\int\Big\{N_+M_++N_-M_-\Big\}dt,
\end{multline}
where 
\begin{equation}\label{1-3}
\begin{split}
\Pi_\Lambda:=&-\frac{M^2_P}{N}R\left(\frac{dR}{dt}-\frac{dR}{dr}N^r\right),\\
\Pi_R:=&-\frac{M_P^2}{N}\Bigg\{\Lambda\left(\frac{dR}{dt}-\frac{dR}{dr}N^r\right)+R\left(\frac{d\Lambda}{dt}-\frac{d}{dr}(\Lambda N^r)\right)\Bigg\},
\end{split}
\end{equation}
are the conjugate momenta of $\Lambda$ and $R$ respectively, $M_P=1/\sqrt{G}$ is the Planck mass, and $\Sigma_r$ denotes the one-dimensional radial Cauchy surface which extended from $-\infty$ to $\infty$. 
Moreover, $H$ and $H_r$ are the super-Hamiltonian and the radial super-momentum constraints, respectively. As it was shown by Kucha\v r \cite{Kuchar1}, solving the Hamiltonian and momentum constraints provides us an observable, namely the mass of BH, $0\leq M<\infty$; correspondingly, the canonical conjugate momenta $-\infty<P_M<\infty$ is given by
\begin{equation}\label{1-4}
P_M:=-\displaystyle\int_{\Sigma_r}\frac{\sqrt{\left(\frac{dR}{dr}\right)^2-\Lambda\left(1-\frac{2M}{M^2_PR}\right)}}{1-\frac{2M}{M^2_PR}}dr.
\end{equation}
This brings the action (\ref{1-2}) to the unconstrained Hamiltonian form
\begin{equation}\label{1-5}
S=\int \Big\{P_M\dot M-(N_++N_-)M\Big\}dt,
\end{equation}
where the new time-dependent conjugate variables $(M,P_M)$ obey the Poisson bracket $\{M,P_M\}=1$. Following Ref. \cite{Louko}, if we chose the right-hand-side asymptotic Minkowski time as observer time parameter, we should restrict to $N_+=1$ and $N_-=0$.  Then, the reduced action (\ref{1-5}) reads as
\begin{equation}\label{1-6}
S=\int \Big\{P_M\dot M-H(M)\Big\}dt,
\end{equation}
where $H(M):=M$ is the reduced Hamiltonian.

The solution of the field equations are $M=\text{const}.$ and $P_M=-t$, as expected. 
The constancy of mass $M$ follows the Birkhoff's theorem,
which states that the mass is the only time-independent and coordinate invariant solution. Furthermore, the conjugate momenta, $P_M$, represents the asymptotic time coordinate at the spacelike slice \cite{Kuchar1}. Since $P_M$ plays the
role of time, it should be periodic in which the period is the inverse Hawking temperature $T_\text{H}=M^2_\text{P}/8\pi M$  \cite{Das,Li}, that is
\begin{equation}\label{1-7}
P_M\sim P_M+\frac{1}{T_\text{H}}.
\end{equation}
The above boundary condition verifies that there is no conical singularity in the $2D$ euclidean section.  Also, it indicates that the phase space is a wedge cut out from the full  $(M, P_M)$ phase space, bounded by the Mass axis and the line $P_M=1/T_\text{H}$ \cite{Med}. Hence, according to the references \cite{Louko,Bar1}, one could 
make the following canonical transformation $(M, P_M ) \rightarrow (x, p_x)$, which  simultaneously opens up the phase space and also incorporates the periodicity condition (\ref{1-7})
\begin{equation}
\begin{split}
x:=\sqrt{\frac{A}{4\pi G}}\cos(2\pi P_MT_\text{H}),\\
p_x:=\sqrt{\frac{A}{4\pi G}}\sin(2\pi P_MT_\text{H}),
\end{split}
\end{equation}
where $A=16\pi M^2/M_\text{P}^4$ is the BH horizon area. From the above canonical transformations, one immediately finds the horizon area in terms of $(x,p_x)$
\begin{equation}\label{1-8}
A=4\pi M_\text{P}^2(x^2+p_x^2).
\end{equation}
This equation shows that the area of the event horizon can be written as the Hamiltonian of a simple harmonic oscillator with the mass $m$ and the angular frequency $\omega$ given by $m=1/\omega=M_\text{P}^2/8\pi $.  In the coordinate representation $\hat p=id/dx$, $\hat x=x$, the canonical quantization is obtained with the time-independent Wheeler--DeWitt equation in one-dimensional minisuperspace
\begin{equation}
4\pi M_\text{P}^2\left(-\frac{d^2}{dx^2}+x^2  \right)\psi(x)=A\psi(x),
\end{equation}
which immediately gives us the area of the event horizon and the mass spectrum \cite{Louko}
\begin{equation}\label{1-10}
A={8\pi L_\text{P}^2}(n+\frac{1}{2}),~~~~
M=\frac{M_\text{P}}{\sqrt{2}}\sqrt{n+\frac{1}{2}},
\end{equation}
where $n$ is an integer.
 Bekenstein \cite{Bec3}  firstly found a similar mass spectrum. Generally, the proportionality constant for the square root of $n$ in (\ref{1-10}) is model dependent.
 Relations (\ref{1-10}) gives the well-known result: Hawking radiation takes place when the BH jumps from a higher state $n+1$ to a lower state $n$, in which the difference in quanta being radiated away. Also, they show that the BH does not evaporate completely, but a Planck size remnant is left over at the end of the evaporation process. In 1974,  Hawking \cite{Hawking} showed that due to quantum fluctuations, BHs emit blackbody radiation and the corresponding entropy is one-fourth of the event horizon area, namely $A=16\pi G^2M^2$.
Following  Refs. \cite{Area1a} and \cite{Xiang}, let us assume that Hawking radiation of a massive BH, i.e., $M \gg M_P$ and  $n\gg 1$, is 
emitted when the BH system  spontaneously jumps from the state $n+1$ into the closest state  level, i.e.,  $n$, as  described by (\ref{1-10}). If we denote the frequency of emitted radiation as $\omega_0$, then
\begin{multline}\label{3-1}
\omega_0=M(n+1)-M(n)\simeq\frac{M_P}{2\sqrt{2n}}\\\simeq\frac{M_P^2}{4M}\left(1+\frac{1}{8}\left(\frac{M_P}{M}\right)^2\right),
\end{multline}
which is agrees with the classical BH oscillation frequencies which scale as $1/M$.
We thus expect a BH to radiate with
a characteristic temperature $T\propto M_P^2/M$, matching the Hawking temperature.

The characteristic BH time scale  (the lifetime of the BH at the state $M(n+1)$ before decaying into the lower state $M(n)$) can be defined \cite{Xiang} as
\begin{equation}\label{3-2}
\tau_n^{-1}:=\frac{|\dot M|}{\omega_0}\simeq\frac{4M|\dot M|}{M_P^2}\left(1-\frac{1}{8}\left(\frac{M_P}{M}\right)^2\right),
\end{equation}
where $\dot M=dM/dt$ is the mass loss of the BH because of its evaporation. Besides, in the second equality, we used the definition of $\omega_0$ expressed in (\ref{3-1}). As discussed in \cite{Area1a,Mukhanov2}, BH interacts with the vacuum of the quantum fields and consequently, the width of states, $W_n$, is not zero. The width of state $n$ can be estimated \cite{Area1a,Mukhanov2} by
\begin{equation}\label{3-3}
|W_n|=\beta (M(n+1)-M(n))=\beta\omega_0,
\end{equation}
where $\beta\ll1$ is a numerical dimensionless factor. Then, by inserting (\ref{3-1}) into the uncertainty relation $W_n\tau_n\simeq 1$ and eliminating $\tau_n$ in resulting equation,  with (\ref{3-2}) we obtain
\begin{equation}\label{3-4}
|\dot M|=\frac{\beta M_P^4}{4M^2}\left(1+\frac{1}{4}\left(\frac{M_P}{M}\right)^2\right).
\end{equation}
If we suppose that the origin of Hawking radiation is the highly blueshifted modes just outside the horizon, and besides, if we assume the BH as a blackbody, then the radiated power is given by the Stefan--Boltzmann law \cite{Radiation,Radiation2}
\begin{equation}\label{3-5}
|\dot M|=\sigma_SAT^4,
\end{equation}
where $\sigma_S=\pi^2/60$ is the Stefan--Boltzmann constant and $A=16\pi M^2/M_P^4$ is the horizon area. Eliminating the mass loss of the BH, using Eqs.(\ref{3-4}) and (\ref{3-5}), gives us the effective temperature 
\begin{equation}\label{3-6}
T=\left(\frac{\beta}{16\pi\sigma_S}\right)^\frac{1}{4}\frac{M_p^2}{4M}\left(1+\frac{1}{16}\left(\frac{M_P}{M}\right)^2\right).
\end{equation}
The BH entropy can be expressed as
\begin{equation}\label{3-7}
S=\int \frac{dM}{T}.
\end{equation}
By choosing $\beta=\sigma_S/\pi^3$, 
Eqs.(\ref{3-6}) and (\ref{3-7}) give us
\begin{equation}\label{3-8}
S=S_\text{BH}-\frac{\pi}{4}\ln{\left(S_\text{BH}\right)}+const.,
\end{equation}
where $S_\text{BH}=4\pi M^2/M_\text{P}^2$ is the  Bekenstein--Hawking entropy. The logarithmic correction to the Bekenstein--Hawking entropy has also been obtained using other methods \cite{Correction2,Correction3,Correction6,Correction7,Correction8,Correction9} except that the overall factor $\pi/4$ is model dependent.
\section{Quantum deformation and the entangled Black Holes-White Holes }
To construct the quantum deformation of the model, let us employ the Heisenberg--Weyl algebra of  $h_4$ which is a non-semisimple Lie algebra with four generators
$\{a_+, a_-, N\}$ that satisfy following commutation relations
\begin{equation}\label{new1}
[a_-,a_+]=1,~~~~[N,a_{\pm}]=\pm a_{\pm},
\end{equation}
where the ladder operators, $\{a_-,a_+\}$, are defined by
\begin{equation}\label{new2}
a_\pm:=\frac{1}{\sqrt{2}}(\pm\frac{d}{dx}+x).
\end{equation}
In the Fock space, $F_2$, with the basis $\{|n\rangle, N|n\rangle = n|n\rangle\}$
the pairs of operators $a_\pm$ act in the following form
\begin{equation}\label{new3}
a_+|n\rangle=\sqrt{n+1}|n\rangle,~~~~a_-|n\rangle
=\sqrt{n}|n\rangle.
\end{equation}
Then the area operator (\ref{1-8}) of the BH will be
\begin{equation}\label{3-12nona}
A=\frac{\omega}{2}\left(a_+a_-+a_- a_+\right).
\end{equation}
The quantum Heisenberg-Weyl algebra, $U_q
(h_4)$, is the
associative unital \cite{Chaichian1} $\mathbb C(q)$-algebra with generators\\
$\{a_+,a_-,q^{\pm N/2}\}$ with the following quantum deformed (q-deformed) commutation relations \cite{Chaichian1}
\begin{equation}\label{new4}
\begin{split}
&a_-a_+-q^{\frac{1}{2}}a_+a_-=q^{\frac{N}{2}},~~~[N,a_\pm]=\pm a_\pm,\\
&a_\pm^\dagger=a_\mp,~~N^\dagger=N,
\end{split}
\end{equation}
where $q\in \mathbb R^+,~~\text{or}~~ q\in \mathbb C~~ \text{and}~~|q|=1$.
Note that in the above definition, we do not postulate any relation among $\{a_\pm,N\}$. One can easily  show
that the first two relations (\ref{new4}) are
actually equivalent to the following relations
\begin{equation}\label{new111}
a_+a_-=[N],~~~a_-a_+=[N+1],
\end{equation}
where 
\begin{equation}\label{number}
[x]:=\frac{q^\frac{x}{2}-q^\frac{-x}{2}}{q^\frac{1}{2}-q^\frac{-1}{2}}.
\end{equation}
One can introduces a linear space of states by defining the vacuum ket $|0\rangle$. The ground state of the quantum deformed Fock space, $ F_2(q)$, is defined by
\begin{equation}\label{3-7non}
a_-|0\rangle=0,\,\,\,\,\,\,\,\,\,q^{\pm N}|0\rangle=|0\rangle.
\end{equation}
Now, like the ordinary Fock space of the harmonic oscillator, we can construct the representation of the $\mathcal U_q(h_4)$ in the q-deformed Fock space spanned by normalized eigenvectors $|n\rangle$
\begin{equation}\label{3-8non}
|n\rangle=\frac{1}{\sqrt{[n]!}}{a}_+^n|0\rangle,
\end{equation}
where the q-factorial defined by $[n]!:=\prod_{m=1}^n[m]$. 
By using the identities (which one can find  from the first relation of (\ref{new4}) by induction)
\begin{equation}\label{3-9non}
a_-a_+^n=q^na_+^na_-+[n]a_+^{n-1}q^{-
N},
\end{equation}
one can show that the basis $|n\rangle$
defined above is orthonormal.
Thus, in the Fock space $\mathcal F_2(q)$, the set up operators act due
\begin{equation}\label{3-10non}
\begin{split}
&a_+|n\rangle=\sqrt{[n+1]}|n+1\rangle,~~~~
a_-|n\rangle=\sqrt{[n]}|n-1\rangle,\\
 &N|n\rangle=n|n\rangle.
\end{split}
\end{equation}
The operator
\begin{equation}\label{3-12non}
A=\frac{\omega}{2}\left(a_+a_-+a_- a_+\right)=\frac{\omega}{2}([N+1]+[N]),
\end{equation}
can be considered to be the $q$-analog of the BH area operator defined in (\ref{3-12nona}). Furthermore, the phase space realization of the $q$-deformed 
phase space pf the BH is given by
\begin{equation}\label{3-13non}
\hat x=\frac{1}{\sqrt{2}}(a_++a_-),~~~~~
\hat p=i{\frac{1}{\sqrt{2}}}( a_+- a_-).
\end{equation}
With  keeping  one  eye  on Eq.(\ref{new111}), the commutation relation of the quantum position  $\hat x$ and the quantum momentum $\hat p$ is not constant anymore 
\begin{equation}\label{3-131non}
[\hat x,\hat p]=i([ N+1]-[ N]),
\end{equation}
which shows that the effective Planck's constant ($\hbar$) is no longer a constant, but depends on the state of the q-deformed BH.

 The preceding content in this section has been based on the assumption that the deformation parameter $q$ is a real and positive $q\in \mathbb R^+$.  If we enlarge $q$ to complex values, then the quantum universal enveloping algebra becomes complex with non-unitary representations. However, in the special case where $q$ is a root of unity, there exist admittedly unitary representations.

Let us now  suppose $q$ is a primitive root of unity, i.e.,
\begin{equation}\label{3-14non}
q=\exp\left(\frac{2\pi i}{\mathfrak{N}}\right),
\end{equation}
where $\mathfrak{N}$ is a natural number, $\mathfrak{N}\in\mathbb{N}^+$, and $\mathfrak{N}\geq2$. 
 Our setting  involves this  natural number, $\mathfrak{N}$, that cause the deformation of mathematical structure of the BH. 

It is clear that for  $\mathfrak{N}\rightarrow\infty$, the deformation parameter
$q\rightarrow1$ and all of the deformed quantities will reduce to the ordinary
undeformed  features. 
The
quantum number defined in (\ref{number}) will be
\begin{equation}\label{3-15non}
[y]=\frac{\sin\left(\frac{\pi y}{\mathfrak{N}}\right)}{\sin\left(\frac{\pi}{\mathfrak{N}}\right)}.
\end{equation}
Now, the action of the ladder operators
$\{a_+, a_-\}$ on the basis eigenvectors are
\begin{equation}\label{3-16non}
\begin{split}
&a_+|n\rangle=\sqrt{[n+1]}|n+1\rangle,~~~~
a_-|n\rangle=\sqrt{[n]}|n-1\rangle,\\
&a_-|0\rangle=0,~~~~a_+|\mathfrak{N}-1\rangle=\sqrt{[\mathfrak{N}]}|\mathfrak{N}\rangle=0,
\end{split}
\end{equation}
where in the last equality we used $[\mathfrak N]=0$.
Therefore, $a_+$  annihilates the state $|\mathfrak{N}-1\rangle$ and the q-deformed Fock space $\mathcal F_2(q)$ is a finite $\mathfrak N$-dimensional vector
space with basis $\{|0\rangle,|1\rangle,...,|\mathfrak{N}-1\rangle\}$. By using identity (\ref{3-9non}) and the defining relations of $\mathcal U_q(h_4)$, Eqs.(\ref{new4}),
on can show that at the root of unity the elements $\{a_+^\mathfrak{N}, a_-^\mathfrak{N}, q^{{\mathfrak{N}} N}, q^{-{\mathfrak{N}} N}\}$ lie in the center of $\mathcal U_q(h_4)$, which means these operators commute with generators of algebra.
The Fock space matrix representation of the generators then become finite-dimensional \cite{1994PhLB..331..150B}
\begin{equation}\label{3-17non}
\begin{split}
&a_+=\displaystyle\sum_{n=0}^{\mathfrak{N}-2}\sqrt{[n+1]}|n+1\rangle\langle n|,\\
&a_-=\displaystyle\sum_{n=1}^{\mathfrak{N}-1}\sqrt{[n]}|n-1\rangle\langle n|,~~~
 N=\displaystyle\sum_{n=0}^{\mathfrak{N}-1}n|n\rangle\langle n|.
 \end{split}
\end{equation}

\begin{figure}[h]
\centering
\includegraphics[width=6cm]{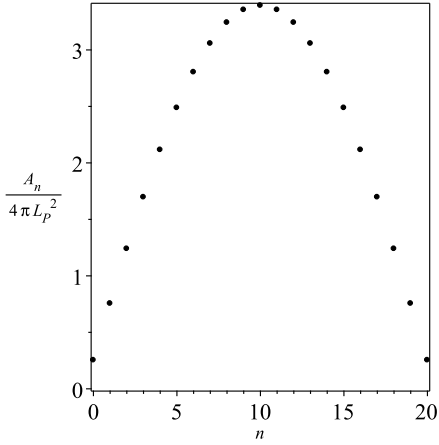}
\caption{Plot of the event horizon of a q-deformed Schwarzschild BH for $\mathfrak N=21$ in units of $4\pi L_\text{P}^2$. }\label{ppp1}
\centering
\end{figure}

Now, Eqs.(\ref{3-12non}) and (\ref{3-15non}) admit the following eigenvalues for the area and the mass of the BH
\begin{equation}\label{3-19non}
A_n={4\pi L_\text{P}^2}\frac{\sin(\frac{\pi}{\mathfrak{N}}(n+\frac{1}{2}))}{\sin(\frac{2\pi}{\mathfrak{N}})},
\end{equation}
\begin{equation}\label{3-19non1}
M_n=\frac{M_\text{P}}{2}\sqrt{\frac{\sin(\frac{\pi}{\mathfrak{N}}(n+\frac{1}{2}))}{\sin(\frac{2\pi}{\mathfrak{N}})}},
\end{equation}
where $n=0,...,\mathfrak{N}-1$.
Note that for $\mathfrak{N}\rightarrow\infty$ the earlier eigenvalues will reduce to
(\ref{1-10}). Since $\sin(\frac{\pi}{\mathfrak{N}}(n+\frac{1}{2}))=\sin(\frac{\pi}{\mathfrak{N}}(\mathfrak{N}-n-1+\frac{1}{2}))$,
there is a two-fold degeneracy at the eigenvalues of the horizon further at the mass spectrum of the BH, see Fig.(\ref{ppp1}). 

To summarize the consequences of the above eigenvalue relations, for simplicity let us consider $\mathfrak N$ is an odd natural number:

1) The area and the mass of the ground state $n=0$ as well as the state $n=\mathfrak N-1$ are
    \begin{equation}
    \begin{split}
        A(n=0~\&~n=\mathfrak N-1)=4\pi L_\text{P}^2,\\
        M({n=0~\&~n=\mathfrak N-1})=\frac{M_\text{P}}{2}. 
    \end{split}
    \end{equation}
    These show that the area and the mass of the ground state are not deformed and their values are the same as the non-deformed spectrums obtained in (\ref{1-10}). Besides, the spectrums are bounded in which the most excited state, $n=\mathfrak N-1$, has the same mass and area as the ground state. In addition, for $n\ll (\mathfrak N-1)/2$ or $(\mathfrak N-1)/2\ll n\leq\mathfrak N-1$, Eqs.(\ref{3-19non}) and (\ref{3-19non1}) will reduce to the non-deformed spectrums obtained in (\ref{1-10}).
    
2) The area and the mass spectrums (\ref{3-19non}) and  (\ref{3-19non1}) represent the spectrums of an Einstein-Rosen wormhole: For $n<(\mathfrak N-1)/2$ it represents a BH which grows by absorbing matter while for $n>(\mathfrak N-1)/2$ it realize a WH (the exact time reversal of the BH) which is exploding and shrinking from a higher state $n=(\mathfrak N-1)/2$ to the ground state $n=0$. The number of possible states for the WH (and also the BH) is finite and is equal to $(\mathfrak N-1)/2$. 
   The upper bound of the area and the mass of the BH are given by
    \begin{equation}\label{mass0}
    \begin{split}
         A(n=\frac{\mathfrak N-1}{2})&=\frac{4\pi L_\text{P}^2}{\sin(\frac{2\pi}{\mathfrak N})},\\
         M(n=\frac{\mathfrak N-1}{2})&=\frac{M_\text{P}}{2\sqrt{\sin(\frac{2\pi}{\mathfrak N})}}.
 \end{split}
        \end{equation}
        
 Consider now the hypothesis that these (primordial) wormholes are a constituent of dark matter.
  Primordial BHs (PBH) are BHs that are thought to have evolved prior to the big bang nucleosynthesis as a result of instability in the early Universe \cite{1979104N,Hawking:1971ei}. Zel'dovich and Novikov \cite{1966AZh758Z} postulated the existence of such BHs in 1966. Chapline \cite{1975Natur51C}, Meszaros \cite{1975AA.5M}, and Hawking and Carr \cite{Hawking:1971ei,1974MNRAS.399C,1975ApJ1C} investigated the formation of such BHs and discovered that, due to their non-relativistic and effectively collisionless nature, they might be viable candidates for dark matter. PBHs have sparked much attention since discovering gravitational waves caused by the merger of massive BHs \cite{LIGO}. 
In the theory of PBH formation, it is a common belief that  BH with the mass $M$  could have formed when its event horizon radius was of the order (or less than) of the CEH.
This idea leads us to consider 
one may consider that the upper bound of the area of the event horizon of the BH is less than the CEH of the Universe, $A_\text{Universe}=4\pi L_\text{CEH}^2$, (where $L_\text{CEH}$ is the CEH radius), at the wormhole formation time, i.e., $A_\text{Max}\leq A_\text{Universe}$.  Saturating this inequality gives us
\begin{equation}\label{mass55}
\mathfrak N=\left(\frac{L_\text{CEH}}{L_\text{P}}\right)^2.
\end{equation}
Generally, the proper distance to the CEH is time dependent. It increase when the Universe is matter or radiation dominated  and remaining constant when the Universe
is dark energy dominated.
Following Gibbons and Hawking, the entropy of the CEH is
\begin{equation}\label{CEH}
S_\text{CEH}=\frac{A_\text{Universe}}{4G}=\pi\left(\frac{L_\text{CEH}}{L_\text{P}}\right)^2.
\end{equation}
Therefore, 
\begin{equation}\label{CEH2}
\mathfrak N=\frac{1}{\pi}S_\text{CEH},~~\text{or}~~~q=\exp({\frac{2\pi^2i}{S_\text{CEH}}}).
\end{equation}

At the Planck epoch after the Big Bang, the scale of the Universe was the Planck length. Hence it is logical to put $\mathfrak N=2$ at the Planck epoch.   For $\mathfrak N=2$ there is only one possible 2-fold degenerate state for the area of horizon, $A=\{4\pi L_\text{P}^2\}$. The mass of this entangled BH/WH configuration is half of the Planck mass. Like the spin of electron (or a qubit), the corresponding Hilbert space is $2D$ with basis $\{|0\rangle,|1\rangle\}$ (or we may use notation $\{|BH\rangle,|WH\rangle\}$), where $|0\rangle$ (or $|BH\rangle$) represents a BH state and $|1\rangle$ (or $|WH\rangle$) is the state of WH. Hence, an arbitrary wavefunction can be expressed as $|\Psi\rangle=c_1|BH\rangle+c_2|WH\rangle$.
A measurement in the $\{|BH\rangle ,|WH\rangle \}$ basis will yield outcome $|BH\rangle$  with probability $|c_1 |^{2}$ and outcome $ |WH\rangle$  with probability $ |c_2 |^{2}$.

Another epoch in which the formation of PBHs is interesting is shortly after reheating. The size of the event horizon of the Universe was presumably in order of
\begin{equation}\label{mass1}
10^{31}L_\text{P}^2\leq A_\text{Universe}\leq 10^{41}L_\text{P}^2.
\end{equation}
For very large values of $\mathfrak N$, Eq.(\ref{mass0}) gives us the upper bound of the mass as \begin{equation}\label{mass2}
M_\text{Max}\simeq 2^{-\frac{3}{2}}\sqrt{\frac{\mathfrak N}{\pi}}M_\text{P}. 
\end{equation}
Eqs.(\ref{mass1}) and (\ref{mass2}) give us the following bound on the value of $\mathfrak N$ just after reheating
\begin{equation}\label{mass4}
10^{15}\leq \mathfrak N\leq 10^{20},
\end{equation}
which shows the number of the quantum states for those wormholes which are formed after reheating. 
To have an idea of the required density of Planck-scale wormholes discussed above, a local dark matter density of the order of $0.01M_\odot/pc^3$ corresponds to approximately two  Planck-scale wormholes per each $10.000~ \text{Km}^3$. Also, the remnant of the wormholes formed after the inflation period (where their quantum number $\mathfrak N$ is given by (\ref{mass4})) could be at the ground state at present. These Planck-scale remnants together with Planck-scale wormholes created after Big Bang could form a component of dark matter at the Universe \cite{Rov1}. The stability of such objects and a possible resolution to the information paradox has been studied in references in \cite{Rov1}.

For the present epoch $t_0$, we obtain
\begin{equation}\label{mass5}
\mathfrak N=\left(\frac{L_\text{CEH}(t_0)}{L_\text{P}}\right)^2\simeq 10^{122},
\end{equation}
where in the last equality we used the value $L_\text{CEH}=15.7\pm 0.4~ Glys\simeq 10^{61} L_\text{P}$ for the CEH radius obtained in Ref. \cite{Egan}.

3)  Let us now calculate the entropy of q-deformed quantum BH. Like to the previous chapter, we assume that Hawking radiation of a massive BH, i.e., $M\gg M_\text{P}$ and $(\mathfrak N-1)/2 \gg n\gg1$, is
emitted when the system spontaneously jumps
from the state $n+ 1$ into the closest state level, i.e., $n$. If we denote the frequency of emitted radiation by $\omega$, then the Mass formula (\ref{3-19non1}) gives us
\begin{multline}\label{qq1}
\omega\simeq\frac{M_P}{2\sqrt{2}}\Bigg\{\frac{1}{\sqrt{n}}-\frac{5}{5}(\frac{\pi}{\mathfrak N})^2n^\frac{3}{2}\Bigg\}
\simeq\\\frac{M_P^2}{4M}\Bigg\{1+\frac{1}{8}\left(\frac{M_P}{M}\right)^2-\frac{10}{3}(\frac{\pi}{\mathfrak N})^2(\frac{M}{M_\text{P}})^4\Bigg\}.
\end{multline}
 Inserting the above frequency into Eq.(\ref{3-2}) gives us the modified 
 life time of the BH at the state $n+1$. If we repeat the same process after Eq.(\ref{3-2}) we find the effective temperature of the BH
 \begin{multline}\label{3-6b}
T=\left(\frac{\beta}{16\pi\sigma_S}\right)^\frac{1}{4}\frac{M_p^2}{4M}\Bigg\{1+\frac{1}{16}\left(\frac{M_P}{M}\right)^2\\-\frac{5}{3}(\frac{\pi}{\mathfrak N})^2(\frac{M}{M_\text{P}})^4\Bigg\}.
\end{multline}
 Inserting this result into (\ref{3-7}) gives us the quantum deformed entropy of massive BHs
 \begin{equation}\label{last}
  S=S_\text{BH}-\frac{\pi}{4}\ln{\left(S_\text{BH}\right)}+\frac{5}{144\pi \mathfrak N^2}S_\text{BH}^3+\text{const}.
 \end{equation}
 Note that for $\mathfrak N\rightarrow\infty$, we will obtain the non-deformed entropy obtained in (\ref{3-8}). In the above equation, one could rewrite the correction terms in terms of the entropy of the CEH. Using Eq.(\ref{CEH2}) we find
 \begin{equation}
 S=S_\text{BH}-\frac{\pi}{4}\ln{\left(S_\text{BH}\right)}+\frac{5\pi}{144S_\text{CEH}^2 }S_\text{BH}^3+\text{const}.
 \end{equation}
 This result shows that at the quantum deformation level, the entropy of the BH contains three parts: the usual Bekenstein--Hawking entropy, the logarithmic term, and a Cube of usual BH entropy.
 The Bekenstein--Hawking entropy the largest observed supermassive BHs (nearly independent of redshift, from the local, $z\simeq0$, to the early, $z > 6$) $S_\text{B}\simeq 10^{104}$ \cite{Egan}. Also, the present entropy of the CEH is $S_\text{CEH}= 2.6\pm0.3\times 10^{122}\simeq10^{122}$. Hence, the CEH contributes almost 18 orders of magnitude more entropy than the supermassive BHs.
  Inserting these estimated value of the entropy into the Eq.(\ref{last}) gives us 
 \begin{equation}
 S\simeq 10^{104}+\frac{26}{\pi}\ln(10)+\frac{5\pi}{144}10^{68},
 \end{equation}
 which shows that our approximation in calculation of entropy is correct for observed supermassive BHs.
 

\section{Conclusions}
 In this paper, we have investigated the quantum deformation of a Schwarzschild BH. The periodic boundary conditions on the conjugate momenta of the mass of the BH leads us to the Heisenberg--Weyl symmetry of the model.   Consequently, the corresponding quantum group of the model after deformation is $U_q(h_4)$. Assuming the q-deformation parameter is a primitive root of unity, we obtained a bounded spectrum for the area of the event horizon of the BH. The attractive feature of this ``quantum deformation of the quantum BH'' is the existence of degenerate states in which they represent a black hole/white hole or wormhole solutions. We show that the dimension of Hilbert space of those wormholes who created just after the Big Bang is two with a mass of half of the Planck mass. These black hole/white hole entangled states could be a good candidate for dark matter.

\section*{Declaration of competing interest}
The author declares that he has no known competing financial interests or personal relationships that could have appeared to influence the work reported in this paper.

\section*{Acknowledgements}
The author would like to thank the anonymous reviewer for his/her insightful suggestions and careful reading of the manuscript.

\bibliographystyle{elsarticle-num}
\bibliography{BH}

\end{document}